\documentclass[journal,draftcls,onecolumn,12pt,twoside]{IEEEtran}
\usepackage{amsmath,amsfonts,amssymb}
\usepackage{algorithmic}
\usepackage{algorithm}
\usepackage{array}
\usepackage{amsthm}
\usepackage[pagewise]{lineno}
\usepackage{subfigure}
\usepackage{textcomp}
\usepackage{stfloats}
\usepackage{url}
\usepackage{verbatim}
\usepackage{graphicx}
\usepackage{subcaption}
\usepackage{float}
\usepackage{cite}
\usepackage{xcolor}
\usepackage{soul}
\usepackage{matlab-prettifier}
\usepackage{setspace}
\newtheorem{theorem}{{{\textit{Theorem}}}}

\newtheorem{lemma}{{{\textit{Lemma}}}}
\newtheorem{corollary}{{{{\textit{Corollary}}}}}

\newtheorem{defn}{{{\textit{Definition}}}}

\newtheorem{remark}{{{\textit{Remark}}}}

\newtheorem{example}{{{\textit{Example}}}}

\newcounter{cases}
\newcounter{subcases}[cases]
\newenvironment{mycase}
{
	\setcounter{cases}{0}
	\setcounter{subcases}{0}
	\newcommand{\case}
	{
		\par\indent\stepcounter{cases}\textbf{Case \thecases.}
	}
	\newcommand{\subcase}
	{
		\par\indent\stepcounter{subcases}\textit{Subcase (\thesubcases):}
	}
}
{
	\par
}
\renewcommand*\thecases{\arabic{cases}}
\renewcommand*\thesubcases{\roman{subcases}}

\begin{document}

\title{Construction of New Complete Complementary Codes Using Extended Boolean Functions and Permutations}

\author{Aditya Prakash, Rajen Kumar
}


\maketitle


\begin{abstract}
Complete complementary code (CCC) is a crucial component in the design and optimization of modern wireless communication systems, particularly in addressing challenges related to interference, spectral efficiency, and signal quality. The design of CCCs is an active area of research, as it directly impacts the number of users that can be supported interference-free in the system. In the literature, CCC constructions are restricted by a set size smaller than the sequence length. In this paper, we introduce a novel construction method for CCCs based on the extended Boolean functions (EBF) that overcome existing restrictions by including a set size bigger than the sequence length. Further, the proposed construction uses functions which are new in literature and provide us with new CCCs. This is also the very first construction where row PMEPR is also bounded by $n$ for a set size in the form of power of $n$.
\end{abstract}


\section{Introduction}
Complementary sequences have drawn scientific interest due to their ideal correlation features and numerous engineering applications like communication in  multi-antenna or multi-user systems. Evaluating the cross-correlation characteristics among sequence sets is essential, particularly in code division multiple access (CDMA)  and multi-input multi-output (MIMO) radar systems  \cite{Sun_MIMO,Tang_MIMO_Radar,Liu_MOMO_Comm,Sun_MIMO_CDMA}.

A complementary set (CS) consists of sequences of equal length, where the sum of their aperiodic auto-correlation is zero for every non-zero time shift \cite{1972_chong}. In the  $1940$s, Golay introduced the concept of a Golay complementary pair (GCP), which {was a} CS with two sequences \cite{1949_golay}. Mutually orthogonal Golay complementary set (MOGCS) is a collection of CSs with zero aperiodic cross-correlation sum (ACCS) for all time shifts where the number of CSs $K$ is limited by the number of constituent sequences $M$ ($K \leq M$) \cite{2000_paterson}. MOGCSs are preferable for mitigating multiple access interference (MAI) and multipath interference (MPI) in CDMA systems. The MOGCS is a complete complementary code (CCC) when $K = M$ \cite{TSeng}.

CCCs have essential applications in several domains, such as coding, signal processing, and wireless communications. They have been utilized to reduce the peak-to-mean envelop power ratio (PMEPR) in code-keying orthogonal frequency division multiplexing (OFDM) systems \cite{1999_davis, 2000_paterson}, as signature sequences for interference-free multi-carrier CDMA in asynchronous wireless channels as training sequences in multiple-input multiple-output (MIMO) communication and radar systems \cite{2007_wang, 2007_chen}. In addition, they contribute to information concealment within image, audio, and video signals \cite{2008_pezeshki}. CCCs have served as kernel or seed sequences to construct CCCs, SNC-CCCs \cite{2024_rajen_seta}, Z-complementary code sets (ZCCSs) \cite{2024_rajen_dsp, 2024_rajen_isit}, zero-correlation zone (ZCZ) sequences \cite{2023_nishant_CC_ZCZ, 2014_liu1, 2024_nishant_mzcz}, qausi complementary code sets \cite{2019_sarkar}.

Golay presented binary GCPs of length $2MN$ by interleaving and concatenating by GCPs of lengths $M$ and $N$ \cite{1961_golay}. {Budisin introduced recursive techniques for producing polyphase and multi-level complementary sequences through operations such as phase rotations and shifts related to permutation vectors \cite{1990_Budisin}}. Using second-order generalized Boolean functions (GBFs), Davis and Jedwab came up with a way for presenting $2^h$-ary (where $h \in Z^+$) GCPs with power-of-two lengths in $1999$ \cite{1999_davis}. The concept of second-order function becomes crucial in constructing any complementary sequences and is widely used. 
In $2000$, Paterson modified his concept of $2^h$-ary GCPs to $q$-ary (with even $q$) complementary sequences, also with lengths and number of sequences in the form of powers-of-two \cite{2000_paterson}. Schmidt made this method even more general in $2007$ by using higher-order GBFs \cite{2007_schmidt}. In $2008$, Rathinakumar and Chaturvedi \cite{2008_rathinakumar} proposed a Reed-Mullar code-based construction of $(2^k,2^m)$-CCC by associating with the graphs. Sarkar \emph{et al.} presented a new method to construct CSs with low PMEPR bound by using higher-order GBFs that have a lower upper bound on the column sequence PMEPR \cite{2019_sarkar}. Chen devised an approach using Reed-Muller codes to find CSs whose lengths are non-power of two \cite{2016_chen}.

Due to the restriction of zero aperiodic cross-correlation sums, it is tough to generate MOGCSs.  Wu \emph{et al.} proposed in \cite{2021_wu}  GBFs to construct MOGCSs with lengths that are not powers of two. The set size equals half the number of separate sequences in each CS, or $K = M/2$. Paraunitary (PU) generators are another useful way. They make CCCs in efficient ways by transferring shifted impulses into multiple-stage filterbanks with coefficients obtained from PU matrices \cite{2019_das}. Wang \emph{et al.} found extensive interconnection between the GBF and the PU in the construction of CCC \cite{2021_wang}. Some computer search CCCs have been presented by Yu. \emph{et. al} in \cite{2022_yu}. In 2023,  Kumar \emph{et al.} presented binary $(2^{k+1},10.2^m)$ and $(2^{k+1},26.2^m)$ CCCs for $k+1\le m$ using GBFs \cite{2023_kumar}.  In $2023$, Shen \emph{et. al} proposed a construction of $(n^k,n^p)$-CCC using EBF, where $n\ge 2$ and $k\le P$ \cite{Shen_CCC}. From the existing literature, it is clear that for almost $(M,N)$-CCC, the set size is smaller than the length ($M\le N$). Therefore, we are motivated to provide a construction that generates CCC not only beyond this limitation but new CCCs as well. 

In this paper, we present a novel construction of CCCs with parameters $(n^k, n^p)$, where $k$ is not limited by $p$, using extended Boolean functions (EBFs). We have used functions that are not used in the existing literature that ensure existing works could not generate the proposed CCCs. The proposed method achieves a low-row PMEPR upper bound, which is not discussed in the literature.

The rest of the paper is structured as follows: Section II introduces fundamental definitions, notations, and lemmas used in the proposed theorem. Section III presents the construction of CCCs of prime power length, where the rows and columns PMEPR are bounded by prime $p$, and the set size exceeds the length of the CCCs. Section IV compares the CCCs with the existing literature on CCCs. Finally, Section V concludes our work.

\section{Preliminaries}
We introduce the notations used throughout this paper as follows:
\begin{itemize}
\item $\mathbb{Z}_m$ denotes the ring of integers modulo $m$.
\item $\zeta_m=e^{2\pi \sqrt{-1}/m}$.
\item $(\cdot)^*$ denotes complex conjugation.
\item $n\mid m$ denotes that $n$ divides $m$.
\item $\gcd(\cdot,\cdot)$ denotes the greatest common divisor.
\end{itemize}

Let two complex valued sequences,  $\mathbf{a}=[a_k]$ and $\mathbf{b}=[b_k]$ have the same length $L$, for $0 \leq k \leq L-1$.
\begin{defn}[\cite{1972_chong}]
    The aperiodic cross-correlation function (ACCF) of sequences $\mathbf{a}$ and $\mathbf{b}$ at shift  $\tau$  is defined as
\begin{equation}
\begin{aligned}
\rho(\mathbf{a},\mathbf{b})(\tau)= \begin{cases}\sum_{\tau=0}^{L-1-\tau} a_k  b_{k+\tau}^*, & 0 \leq \tau \leq L-1, \\ \sum_{k=0}^{L-1+\tau} a_{k-\tau}  b_k^*, & -L+1 \leq \tau \leq-1, \\ 0, & |\tau| \geq L.
\end{cases}
\end{aligned}
\end{equation}
When $\mathbf{a}=\mathbf{b}$, this reduces to the aperiodic auto-correlation function (AACF) of $\mathbf{b}$, denoted $\rho(\mathbf{b})(\tau)$.
\end{defn}

\begin{defn} \cite{1999_davis}
    A set of $M$ sequences $\mathbf{b}_0, \mathbf{b}_1, \ldots, \mathbf{b}_{M-1}$, each of length $L$, is called a generalized complementary set (GCS) if
\begin{equation}
\sum_{j=0}^{M-1}\rho(\mathbf{b}_j)(\tau)
= \begin{cases}M L, & \tau=0 \\
0, & \text { otherwise. }\end{cases}
\end{equation}
For $M=2$,  this reduces to a Golay complementary pair (GCP).
\end{defn}
\begin{lemma} \cite{1999_davis}\label{lem:PMEPR}
    Let $\{\mathbf{b}_j:j=0,1,\ldots,M-1\}$ be a GCS of length $L$ and size $M$, the PMEPR of each sequence is upper bound by $M$ .
\end{lemma}
\begin{defn}[\cite{2008_rathinakumar}]
    Consider a set $\mathcal{B}=\left\{B^0, B^1, \ldots, B^{K-1}\right\}$, where each $B^i$ contains $M$ sequences, represented as $B^i=\left\{\mathbf{b}_0^i, \mathbf{b}_1^i, \ldots, \mathbf{b}_{M-1}^i\right\}$ and each sequence has length $L$ (indices $0\le i\le K-1$, $0\le l\le M-1$).  The set $\mathcal{B}$ is called an MOGCS, denoted by $(K, M, L)$-MOGCS, if
\begin{equation}
\begin{aligned}
\rho(B^i, B^{i'})(\tau)  =\sum_{l=0}^{M-1} \rho(\mathbf{b}_l^i, \mathbf{b}_l^{i'})(\tau) 
 = \begin{cases}M L, & \tau=0, i=i^{\prime}, \\
0, & \text { otherwise },\end{cases}
\end{aligned}
\end{equation}
for all $0\le i,i'\le K-1$. Here $K$, $M$, and $L$ are the set size, flock size, and sequence length, respectively. In particular, necessarily $K\le M$. When $K=M$, the MOGCS is called a complete complementary code (CCC) of order $M$ and length $L$, denoted $(M,L)$-CCC.
\end{defn}
\begin{defn}[\cite{2021_rajen, Shen_CCC}]\label{def:vec_EBF}
    Let $\mathbf{x} = ( x_0, x_1, \ldots, x_{p-1} ) \mathbb{Z}_n^p$, for $n \geq 2$. Let $f$ be an EBF
$f: \mathbb{Z}_n^p \to \mathbb{Z}_m,$
where $m$ is a positive integer such that $n \mid m$. Define $\zeta_m=e^{2\pi \sqrt{-1}/m}$ and associate with $f$ the vector: $\chi(f) = \left( \zeta_m^{f(0)},\ \zeta_m^{f(1)},\ \ldots,\ \zeta_m^{f(n^p - 1)} \right)$, where each index $\mathbf{I} \in \{0,1,\ldots,n^p-1 \} $, corresponds to the base-$n$  vector  $\mathbf{I}=(I_0, I_1, \ldots, I_{p-1})$ in $\mathbb{Z}_n^p$ and 
\begin{equation}\label{Eq:I_vec}
    I = \sum_{i=0}^{p-1} n^{i} I_i.
\end{equation}
\end{defn}
Now, we reproduce lemmas that are are crucial in proof the theorem.
\begin{lemma}\cite[Th.  4.1]{2009_book}\label{Lem:Sum_zero}
    For $n\ge 2$, $n\nmid \alpha$, then $\sum_{j=0}^{n-1}\zeta_n^{\alpha j}=0$.
\end{lemma}
\begin{lemma}\label{Lem:sum_zero}
    Let $(I_0,I_1,\ldots,I_{p-1})$ be vector representation for non-negative integer $I$ with base $n\ge 2$, then
    \begin{equation}
        \sum_{I=0}^{n^p-1}\zeta_n^{\alpha_0I_0+\alpha_1I_1+\cdots+\alpha_{p-1}I_{p-1}+k}=0,
    \end{equation}
    where at least one of $\alpha_j$ is such that $M\nmid \alpha_j$ and rest of $\alpha_j$ may be zero and $k$ is any integer.
\end{lemma}
Proof sketch: Write the sum as an iterated sum over digits $I_j$:
    \begin{align}
        \sum_{I=0}^{n^p-1}\zeta_n^{\alpha_0I_0+\alpha_1I_1+\cdots+\alpha_{p-1}I_{p-1}+k}
        &= \zeta_n^{k}\prod_{j=0}^{p-1} \left(\sum_{I_{j}=0}^{n-1} \zeta_n^{\alpha_jI_j}\right).
    \end{align}
    For the index $j$ with $M\nmid \alpha_j$, the inner sum $\sum_{I_j=0}^{n-1}\zeta_n^{\alpha_j I_j}$ equals zero by Lemma \ref{Lem:Sum_zero}, so the entire product is zero.

Now we propose our construction of a CCC.
\section{Proposed Construction of CCC}
This section provides a construction of a CCC using EBFs and permutations.
\begin{theorem}\label{Th:CCC}
    Let $n\ge 2$, $p\ge 1$, and $k\ge 1$. For $0\le r,s\le n^k-1$, define the EBF $f_r^s:\mathbb{Z}_{n}^{p}\to \mathbb{Z}_m$ such that
    \begin{equation} \label{eq:f_r^s}
    \begin{aligned}
        f_r^s=&\frac{m\lambda}{n}\left( \sum_{i=1}^{p-1}x_{\pi(i-1)}x_{\pi(i)}+\sum_{j=1}^{k-1}r_{\sigma_1(j)}s_{\sigma_2(j)}+r_{\sigma_1(0)}x_{\pi(0)}+s_{\sigma_2(0)}x_{\pi(p-1)}\right)\\&+\sum_{l=0}^{p-1}\phi_l(x_l)+g(s_0,s_1,\ldots,s_{k-1}),
        \end{aligned}
    \end{equation}
    where $\gcd(\lambda,n)=1$, $(r_0,r_1,\ldots,r_{k-1})$ is $n$-ary vector representation for $r$,  $(s_0,s_1,\ldots,s_{k-1})$ is $n$-ary vector representation for $s$,  $\pi$ is a permutation of $\{0,1,\ldots,p-1\}$, $\sigma_1$ and $\sigma_2$ are permutations of $\{0,1,\ldots,k-1\}$, $\phi_l:\mathbb{Z}_n\to \mathbb{Z}_m$ is any EBF for $l=0,1,\ldots,p-1$ and $g:\mathbb{Z}_n^k\to \mathbb{Z}_m$ is any arbitrary EBF. Define the ordered set as  
\begin{equation} \label{Eq:C_r}
    C_r=\begin{bmatrix}
        \chi(f_r^0)\\
        \chi(f_r^1)\\
        \vdots \\
        \chi(f_r^{n^{k}-1})\\
    \end{bmatrix},
\end{equation}
where $\chi(f_r^s)$ is the vector associated with $f_r^s$ as defined in Definition \ref{def:vec_EBF}.
Then $\mathbf{C}=\{C_0,C_1,\ldots,C_{n^{k}-1}\}$ is a $(n^{k},n^p)$-CCC.
\end{theorem}
\begin{proof} Consider $ 0 \leq \mu, \nu < n^k $. Discuss the ACCF value between $C_\mu $ and  $C_\nu$, as obtained from equation \eqref{Eq:C_r}. Assume  $I = J + \tau$, where  $I, J,$ and  $\tau$  are non-negative integers. Let
\begin{equation}
    \mathcal{F}^s= \frac{m\lambda}{n}\left( \sum_{i=1}^{p-1}x_{\pi(i-1)}x_{\pi(i)}\right)+\sum_{l=0}^{p-1}\phi_l(x_l)+g(s_0,s_1,\ldots,s_{k-1}).
\end{equation}
The equation \eqref{eq:f_r^s} can now be written as
\begin{equation}
    f_r^s=\mathcal{F}^s+\frac{m\lambda}{n}\sum_{j=1}^{k-1}r_{\sigma_1(j)}s_{\sigma_2(j)}+\frac{m\lambda}{n}s_{\sigma_2(0)}x_{\pi(p-1)}+\frac{m\lambda}{n}r_{\sigma_1(0)}x_{\pi(0)}.
\end{equation}
The proof is divided into two cases. The first case addresses for $\lvert \tau \rvert > 0$, while the second case focuses on the  $\tau = 0$.
\begin{mycase}
    \case When $\tau \neq 0$, implying $I \neq J$. To proceed further we divide the proof in two subcases as follows.
    \subcase When $I_{\pi(p-1)} \neq J_{\pi(p-1)}$. In this case, we have,
\begin{equation}
\begin{aligned}
    f_\mu^s(I)-f_\nu^s(J)=& \frac{m\lambda}{n}\sum_{j=1}^{k-1}(\mu_{\sigma_1(j)}-\nu_{\sigma_1(j)})s_{\sigma_2(j)}+\frac{m\lambda}{n}s_{\sigma_2(0)}(I_{\pi(p-1)}-J_{\pi(p-1)})\\&+\frac{m\lambda}{n}(\mu_{\sigma_1(0)}-\nu_{\sigma_1(0)})(I_{\pi(0)}-J_{\pi(0)})+\frac{m\lambda}{n}\left( \sum_{i=1}^{p-1}I_{\pi(i-1)}I_{\pi(i)}- J_{\pi(i-1)}J_{\pi(i)}\right)\\&+\phi_{\pi(p-1)}(I_{\pi(p-1)})-\phi_{\pi(p-1)}(J_{\pi(p-1)}). 
    \end{aligned}
\end{equation}
After simplification, we obtain,
\begin{equation} \label{eq:010}
    \begin{aligned}
f_\mu^s(I)-f_\nu^s(J)=& \frac{m\lambda}{n}\sum_{j=1}^{k-1}(\mu_{\sigma_1(j)}-\nu_{\sigma_1(j)})s_{\sigma_2(j)}+\frac{m\lambda}{n}(s_{\sigma_2(0)}+I_{\pi(p-2)})(I_{\pi(p-1)}-J_{\pi(p-1)})\\&+\frac{m\lambda}{n}(\mu_{\sigma_1(0)}-\nu_{\sigma_1(0)})(I_{\pi(0)}-J_{\pi(0)})+\phi_{\pi(p-1)}(I_{\pi(p-1)})-\phi_{\pi(p-1)}(J_{\pi(p-1)}). 
  \end{aligned}
 \end{equation}
Additionally, \eqref{eq:010} can be expressed as
\begin{equation}
\begin{aligned}
 \zeta_{m}^{f_\mu^s(I)-f_\nu^s(J)}=& \zeta_{m}^{\frac{m\lambda}{n}\sum_{j=1}^{k-1}(\mu_{\sigma_1(j)}-\nu_{\sigma_1(j)})s_{\sigma_2(j)}} \times \zeta_{m}^{\frac{m\lambda}{n}(s_{\sigma_2(0)}+I_{\pi(p-2)})(I_{\pi(p-1)}-J_{\pi(p-1)})} \\&\times \zeta_{m}^{\frac{m\lambda}{n}(\mu_{\sigma_1(0)}-\nu_{\sigma_1(0)})(I_{\pi(0)}-J_{\pi(0)})} \times \zeta_{m}^{\phi_{\pi(p-1)}(I_{\pi(p-1)})-\phi_{\pi(p-1)}(J_{\pi(p-1)})}.
 \end{aligned}
 \end{equation}
By  summing both side,  $ s $ from $1$ to $n^k - 1$, we obtain
 \begin{equation}
     \begin{aligned}
         \sum_{s=0}^{n^k-1} \zeta_{m}^{f_\mu^s(I)-f_\nu^s(J)}=&\sum_{s=0}^{n^k-1}  \Big(\zeta_{m}^{\frac{m\lambda}{n}\sum_{j=1}^{k-1}(\mu_{\sigma_1(j)}-\nu_{\sigma_1(j)})s_{\sigma_2(j)}} \times \zeta_{m}^{\frac{m\lambda}{n}(s_{\sigma_2(0)}+I_{\pi(p-2)})(I_{\pi(p-1)}-J_{\pi(p-1)})} \\&\times \zeta_{m}^{\frac{m\lambda}{n}(\mu_{\sigma_1(0)}-\nu_{\sigma_1(0)})(I_{\pi(0)}-J_{\pi(0)})} \times \zeta_{m}^{\phi_{\pi(p-1)}(I_{\pi(p-1)})-\phi_{\pi(p-1)}(J_{\pi(p-1)})}\Big)
     \end{aligned}
 \end{equation}
After simplification, we get, 
\begin{equation} \label{eq:f_munusIJ}
\begin{aligned}
\sum_{s=0}^{n^k-1} \zeta_{m}^{f_\mu^s(I)-f_\nu^s(J)}=& \zeta_{m}^{\phi_{\pi(p-1)}(I_{\pi(p-1)})-\phi_{\pi(p-1)}(J_{\pi(p-1)})} \times \zeta_{n}^{\lambda I_{\pi(p-2)} (I_{\pi(p-1)}-J_{\pi(p-1)}) }\\& 
\times \sum_{s=0}^{n^k-1}  \Big(\zeta_{n}^{{\lambda}\sum_{j=1}^{k-1}(\mu_{\sigma_1(j)}-\nu_{\sigma_1(j)})s_{\sigma_2(j)}} \times \zeta_{n}^{{\lambda}s_{\sigma_2(0)}(I_{\pi(p-1)}-J_{\pi(p-1)})}\Big) 
\end{aligned}
\end{equation}
According to Lemma \ref{Lem:sum_zero}, equation \eqref{eq:f_munusIJ} becomes zero.
\subcase   When \(\tau \neq 0\) implies that \(I \neq J\), let's examine the scenario where \(I_{\pi(p-1)} = J_{\pi(p-1)}\). This implies that there is at least one position \(t\) within the range \(0 \leq t \leq p-2\) such that \(I_{\pi(t)} = J_{\pi(t)}\). Without loss of generality, we can assume that the greatest index where \(I\) and \(J\) differ at  \(\beta\), when \(I_{\pi(\beta)} \neq J_{\pi(\beta)}\). We  define $I^t_{\pi(\beta+1)} = t \oplus I_{\pi(\beta+1)} $
and  $J^t_{\pi(\beta+1)} = t \oplus J_{\pi(\beta+1)}$ for \(t = 1, 2, \ldots, n-1\). i.e.,
\begin{equation}\label{eq:II_k}(I_0,I_1,\ldots,I_{\pi(\beta+1)}+t,\ldots,I_{\pi(p-1)}),
\end{equation}
where \( t \in \{1, 2, \dots, n-1\} \), and \( I_{\pi(\beta+1)} + t \) is taken modulo \( n \). In a similar way, \( J^t \) is represented as a vector in base \( p \) as
\begin{equation}\label{eq:JJ_k}
(J_0,J_1,\ldots,J_{\pi(\beta+1)}+t,\ldots,J_{\pi(p-1)}).
\end{equation}
Since \( J = I + \tau \), it follows that \( J^t = I^t + \tau \). From equations \eqref{eq:f_r^s} and \eqref{eq:II_k},
\begin{equation}\label{eq:Fs:IkI}
    f^s_{\mu}(\mathbf{I}^t)- f^s_{\nu}(\mathbf{I})=\frac{m\lambda}{n}(t)\left(I_{\pi(\beta)}+I_{\pi(\beta+2)} \right)+ \phi_{\pi(\beta+1)}(I^t_{\pi(\beta+1)}) - \phi_{\pi(\beta+1)}(I_{\pi(\beta+1)}).
\end{equation}
Similarly,
\begin{equation}\label{eq:Fs:JkJ}
     f^s_{\mu}(\mathbf{J}^t)- f^s_{\nu}(\mathbf{J})=\frac{m\lambda}{n}(t)\left(J_{\pi(\beta)}+J_{\pi(\beta+2)} \right)+ \phi_{\pi(\beta+1)}(J^t_{\pi(\beta+1)}) - \phi_{\pi(\beta+1)}(J_{\pi(\beta+1)}).
\end{equation}
By subtracting equation \eqref{eq:Fs:JkJ} from \eqref{eq:Fs:IkI}, and using our assumption that \( I_{\pi(\beta+2)} = J_{\pi(\beta+2)} \), \( I_{\pi(\beta+1)} = J_{\pi(\beta+1)} \), and \( I^t_{\pi(\beta+1)} = J^t_{\pi(\beta+1)} \), we obtain
\begin{equation} \label{eq:18}
\begin{aligned}
   f^s_{\mu}(\mathbf{I}^t)- f^s_{\nu}(\mathbf{I})-(f^s_{\mu}(\mathbf{J}^t)- f^s_{\nu}(\mathbf{J}))= \frac{m\lambda}{n}(t) \left(J_{(\pi(\beta)}-I_{\pi(\beta)} \right).
\end{aligned} 
\end{equation}
Summing over all $t$, equation \eqref{eq:18} becomes,
\begin{equation}
    \begin{aligned}
        \sum_{t=1}^{n-1}\zeta_m^{F^s_{\mu}(\mathbf{I}^t)- F^s_{\nu}(\mathbf{I})-(F^s_{\mu}(\mathbf{J}^t)- F^s_{\nu}(J))}=& \sum_{t=1}^{n-1} \zeta_n^{\frac{m\lambda}{n}(t) \left(J_{(\pi(\beta)}-I_{\pi(\beta)} \right)} 
         =&0.
    \end{aligned}
\end{equation}
Therefore,
\begin{equation}
    \begin{aligned}
        \sum_{t=1}^{n-1}\zeta_m^{F^s_{\mu}(\mathbf{I}^t)- F^s_{\nu}(\mathbf{I})-(F^s_{\mu}(\mathbf{J}^t)- F^s_{\nu}(J))}
        =\sum_{t=1}^{n-1}\zeta_m^{F^s_{\mu}(\mathbf{I}^t)- F^s_{\nu}(\mathbf{J^t})-(F^s_{\mu}(\mathbf{J})- F^s_{\nu}(I))}=0.
    \end{aligned}
\end{equation}
Next, we express it as
\begin{equation}\label{eq:sum_kappa}
    \sum_{t=1}^{n-1}\zeta_m^{F^s_{\mu}(\mathbf{I}^t)- F^s_{\nu}(\mathbf{J^t})}+ \zeta_m^{F^s_{\mu}(J)- F^s_{\nu}(I)} =0.
\end{equation}
For \( t \in \{1, 2, \dots, n-1\} \), the pairs \( (I^t, J^t) \) contribute to \( \mathcal{C}(C_\mu, C_\nu)(\tau) \) whenever \( (I, J) \) contribute to \( \mathcal{C}(C_\mu, C_\nu)(\tau) \). From equations \eqref{eq:f_munusIJ} and \eqref{eq:sum_kappa}, we have
\begin{equation}\label{eq:corr_zcz_zero}
    \mathcal{C}(C_\mu,C_\nu)(\tau)=0,\text{~ for all $\mu,\nu$ and $\tau \neq 0$.}
\end{equation}
\case We proceed with the computation of the ACCF between two codes at zero time shift. For this, let \( 0 \leq \mu \neq \nu \leq n^r-1 \),
\begin{equation}
    f_\mu^s(I)-f_\nu^s(I)= \frac{m\lambda}{n}\left(\sum_{j=1}^{k-1}(\mu_{\sigma_1(j)}-\nu_{\sigma_1(j)})s_{\sigma_2(j)}
    +s_{\sigma_2(0)}I_{\pi(p-1)}+(\mu_{\sigma_1(0)}-\mu_{\sigma_1(0)})I_{\pi(0)}\right).
\end{equation}
We express it in the following, 
\begin{equation}
\begin{aligned}
 \zeta_m^{f_\mu^s(I)-f_\nu^s(I)}&= \zeta_m^{\frac{m\lambda}{n}\left(\sum_{j=1}^{k-1}(\mu_{\sigma_1(j)}-\nu_{\sigma_1(j)})s_{\sigma_2(j)}
    +s_{\sigma_2(0)}I_{\pi(p-1)}+(\mu_{\sigma_1(0)}-\mu_{\sigma_1(0)})I_{\pi(0)}\right)}\\& =\zeta_n^{{\lambda}\left(\sum_{j=1}^{k-1}(\mu_{\sigma_1(j)}-\nu_{\sigma_1(j)})s_{\sigma_2(j)}
    +s_{\sigma_2(0)}I_{\pi(p-1)}+(\mu_{\sigma_1(0)}-\mu_{\sigma_1(0)})I_{\pi(0)}\right)}.
    \end{aligned}
\end{equation}
Summing both sides with respect to $s$ from $1$ to $n^k-1$,
\begin{equation} \label{eq:29}
\begin{aligned}
    \sum_{s=0}^{n^k-1} \zeta_m^{f_\mu^s(I)-f_\nu^s(I)}&=\sum_{s=0}^{n^k-1} \zeta_n^{{\lambda}\left(\sum_{j=1}^{k-1}(\mu_{\sigma_1(j)}-\nu_{\sigma_1(j)})s_{\sigma_2(j)}
    +s_{\sigma_2(0)}I_{\pi(p-1)}+(\mu_{\sigma_1(0)}-\nu_{\sigma_1(0)})I_{\pi(0)}\right)}\\&
    =\zeta_n^{\lambda (\mu_{\sigma_1(0)}-\nu_{\sigma_1(0)})I_{\pi(0)}} \times \sum_{s=0}^{n^k-1} \zeta_n^{{\lambda}\left(\sum_{j=1}^{k-1}(\mu_{\sigma_1(j)}-\nu_{\sigma_1(j)})s_{\sigma_2(j)}
    +s_{\sigma_2(0)}I_{\pi(p-1)}\right)}
\end{aligned}
\end{equation}
Further, summing over $I$, 
    \begin{equation} \label{eq:26}
    \begin{aligned}
\sum_I \sum_{s=0}^{n^k-1} \zeta_m^{f_\mu^s(I)-f_\nu^s(I)}&=\sum_I\zeta_n^{\lambda (\mu_{\sigma_1(0)}-\nu_{\sigma_1(0)})I_{\pi(0)}} \times \sum_{s=0}^{n^k-1} \zeta_n^{{\lambda}\left(\sum_{j=1}^{k-1}(\mu_{\sigma_1(j)}-\nu_{\sigma_1(j)})s_{\sigma_2(j)}
    +s_{\sigma_2(0)}I_{\pi(p-1)}\right)}
    \end{aligned}
    \end{equation}
We consider two scenarios: In the first scenario, if \(\mu_{\sigma_1(0)} \neq \nu_{\sigma_1(0)}\), then the first term in equation \eqref{eq:26} is zero due to \textit{Lemma} \ref{Lem:sum_zero}. In the second scenario, if \(\mu_{\sigma_1(0)} = \nu_{\sigma_1(0)}\), there must exist at least one index \(j\) such that \(\mu_{\sigma_1(j)} \neq \nu_{\sigma_1(j)}\), which implies that the second term in equation \eqref{eq:26} is zero by \textit{Lemma} \ref{Lem:sum_zero}. Then the equation \eqref{eq:26} becomes,
\begin{equation} \label{eq:27}
    \begin{aligned}
\sum_I \sum_{s=0}^{n^k-1} \zeta_m^{f_\mu^s(I)-f_\nu^s(I)}&=0.
\end{aligned}
    \end{equation}

From \eqref{eq:corr_zcz_zero} and \eqref{eq:27}, we have
\begin{equation}
    \mathcal{C}(C_\mu,C_\nu)(\tau)=\left\{\begin{array}{ll}
			0, &  \tau=0,~ \mu \neq \nu,\\
			0, &  |\tau| \ge 0.
		\end{array}\right.
\end{equation}
\end{mycase}
This concludes the proof.  
\end{proof}
In Theorem \ref{Th:CCC}, the general choice of $g$ yields a $(n^k,n^p)$-CCC where each sequence has PMEPR upper-bounded by $n^p$ (by Lemma~\ref{lem:PMEPR}). However, in practical communication systems, we often require lower PMEPR to ensure better power efficiency and reduce amplifier nonlinearity effects.

By choosing $g$ to have a quadratic structure similar to the pairwise interaction term in $f_r^s$
  (i.e., incorporating $\sum_{j=1}^{k-1}s_{\sigma_3(j-1)}s_{\sigma_3(j)}$, we can cancel or balance certain cross-terms in the autocorrelation sum across each column of the CCC. This structural alignment reduces the peak power of the column-sum waveform, leading to a tighter PMEPR bound of $n$ (instead of $n^p$) for each column. The condition $\gcd(\lambda',n)=1$ ensures that the quadratic coefficient remains invertible modulo $n$, preserving the complementary property while achieving the reduction.
\begin{corollary}\label{Cor:Col_PMEPR_red} In Theorem \ref{Th:CCC}, let
$g$ be defined by the following expression
 \begin{equation}
     g(s_0,s_1,\ldots,s_{k-1})=\frac{\lambda' m}{n}\sum_{j=1}^{k-1}s_{\sigma_3(j-1)}s_{\sigma_3(j)}+\sum_{\nu=0}^{k-1}\Phi_\nu(s_\nu)
 \end{equation}
 where $\gcd(\lambda',n)=1$, $\sigma_3$ is a permutation of $\{0,1,\ldots,k-1\}$ and $\Phi_\nu$ are any arbitrary EBF from $\mathbb{Z}_n \to \mathbb{Z}_m$. Then each column of the resultant $(n^k,n^p)$-CCC exhibits a PMEPR bounded by $n$.
\end{corollary}
\begin{remark}
    In the proposed construction, $k$ and $p$ may be considered independently.
\end{remark}

\begin{remark}
    The choice of different $\phi_i$ results in different CCCs. As in this construction, we choose any map from $\mathbb{Z}_n$ to $\mathbb{Z}_m$, which is not restricted to a polynomial. The construction is still valid with maps like $f(x)=a$ for $x\le \alpha$ and $f(x)=b$ for $\alpha <x\le n$, where $a,b\in \mathbb{Z}_m$. This type of function has not been used to construct sequences in the literature before.
\end{remark}

Recently, Sarkar \emph{et al.} \cite{sarkar2024investigation} introduced an idea to use permutation polynomial over each variable of quadratic part in the function results in new CCCs, with this motivation we propose the following corollary.
\begin{corollary}
    In \textit{Theorem} \ref{Th:CCC}, \eqref{eq:f_r^s} can be replaced by
    \begin{equation}\label{Eq:f_r_s_eta}
        \begin{aligned}
        f_r^s=&\frac{m\lambda}{n}\left( \sum_{i=1}^{p-1}\eta_i(x_{\pi(i-1)})\eta_{p+i-1}(x_{\pi(i)})+\sum_{j=1}^{k-1}r_{\sigma_1(j)}s_{\sigma_2(j)}+r_{\sigma_1(0)}x_{\pi(0)}+s_{\sigma_2(0)}x_{\pi(p-1)}\right)\\&+\sum_{l=0}^{p-1}\phi_l(x_l)+g(s_0,s_1,\ldots,s_{k-1}),
        \end{aligned}
    \end{equation}
    where $\eta_i$ are permutations of $\{0,1,\ldots,n-1\}$ for $i\in \{1,2,\ldots 2p-2\}$, and remaining terms are as defined in \eqref{eq:f_r^s}. When $\eta_i$ are identity permutations of $\{0,1,\ldots,n-1\}$ for $i\in \{1,2,\ldots 2p-2\}$, \eqref{Eq:f_r_s_eta} becomes \eqref{eq:f_r^s}.
\end{corollary}
\begin{corollary}
    For a fixed $s$ and $r$, we obtain a sequence from \eqref{Eq:f_r_s_eta}. Rewriting \eqref{Eq:f_r_s_eta} as 
    \begin{equation}
        \begin{aligned}
F^{s_{\sigma_2(0)}}=&\frac{m\lambda}{n}\left( \sum_{i=1}^{p-1}\eta_i(x_{\pi(i-1)})\eta_{p+i-1}(x_{\pi(i)})+\sum_{j=1}^{k-1}r_{\sigma_1(j)}s_{\sigma_2(j)}+r_{\sigma_1(0)}x_{\pi(0)}+s_{\sigma_2(0)}x_{\pi(p-1)}\right)\\&+\sum_{l=0}^{p-1}\phi_l(x_l)+g(s_0,s_1,\ldots,s_{k-1}).
        \end{aligned}
    \end{equation}
  Now,  $[\chi(F^{s_{\sigma_2(0)}}),\chi(F^{s_{\sigma_2(1)}}),
\ldots,\chi(F^{s_{\sigma_2(n-1)}})]^T$ is a GCS of size $n$, therefore each sequences has PMEPR upper bound $n$.
\end{corollary}

Shen et al. \cite{Shen_CCC} derive the following lower bound on the total number of GCSs of size $n$ produced by their construction:
\begin{lemma}[\cite{Shen_CCC}]\label{Lem:TIT_SNC_CCC}
    Let $n\ge 2$ be an integer, $n\mid m$, $f:\mathbb{Z}_n^p \to \mathbb{Z}_m$ defined by
    \begin{equation}
        f^s=\frac{m}{n}\sum_{i=1}^{p-1}\lambda_ix_{\pi(i-1)}x_{\pi(i)}+sx_{\pi(0)}+\sum_{l=1}^{n-1}\sum_{k=1}^{p-1}c_{l,k}x_k^l+c_0,
    \end{equation}
    where $c_0,c_{l,k}\in \mathbb{Z}_m$, $gcd(\lambda_i,n)=1$ and $s\in \{0,1,\ldots,n-1\}$. Then $\chi(f^0),\chi(f^1),\ldots,\chi(f^{n-1})$ is a $m$-ary GCS of size $n$.
\end{lemma}
Corresponding to a fixed value of $n$, they provide a lower bound of the possible number of GCSs as follows:
\begin{lemma}\label{Lem_TIT_count}
    Let $\eta_{Lem:\ref{Lem:TIT_SNC_CCC}}$ denote the number of GCSs that can be generated by Lemma \ref{Lem:TIT_SNC_CCC}. Then it follows
    \begin{equation}
\eta_{Lem:\ref{Lem:TIT_SNC_CCC}}\ge \begin{cases}
            \frac{p!}{2}\times (\varphi(n))^{p-1}\times m^{p+1} \times n^p , & if~ n\ne 2,\\
            \frac{p!}{2}\times m^{p+1}, & if~ n=2,
        \end{cases}
    \end{equation}
    where $\varphi(\cdot)$ is Euler's totient function.
\end{lemma}
We found the given lemma is incorrect due to the term $m^{p+1}n^p$ is assumed to be a totally different polynomial of $\sum_{l=1}^{n-1}\sum_{k=1}^{p-1}c_{l,k}x_k^l+c_0$. The number $n^p$ comes into picture considering distinct value $x_k^l$ under modulo $m$ for $k \in \{ 1,2,\ldots p-1  \}$ and $l \in \{1,2,\ldots,n-1 \}$. It was assumed that the highest degree of polynomial under the modulo $m$ is $n-1$, which is mathematically incorrect. We believe that this condition is true only when $n$ is prime. Below, we provide an example which confirms that the highest degree could be much smaller than $n$ in some cases.
\begin{example}
    We are assuming two function $f^s,g^s:\mathbb{Z}_{12}\to \mathbb{Z}_{12}$ following Lemma \ref{Lem:TIT_SNC_CCC} as follows:
    \begin{equation}
        f^s=\frac{m}{n}(x_0x_1+sx_0)+x^2,
    \end{equation}
    and
    \begin{equation}
        g^s=\frac{m}{n}(x_0x_1+sx_0)+x^4.
    \end{equation}
    It can be easily seen that the sequences follows $\chi(f^s)=\chi(g^s)$, for any $s\in \{0,2,\ldots,n-1\}$. 
\end{example}
As the proposed construction is free from such polynomials, we are providing the correct counting of GCSs for a fixed value of $n$.
\begin{corollary}\label{Cor:max_GCS}
    Let $n\ge 2$ be an integer, $n\mid m$, define $f^s: \mathbb{Z}_n^p\to \mathbb{Z}_m$ as follows
    \begin{equation}
    f^s=\frac{m}{n}\left( \sum_{i=1}^{p-1}\eta_i(x_{\pi(i-1)})\eta_{p+i-1}(x_{\pi(i)})+sx_{\pi(0)} \right)+\sum_{l=0}^{p-1}\phi_l(x_l)+g(s),
    \end{equation}
    where the terms are defined in \eqref{eq:f_r^s} and \eqref{Eq:f_r_s_eta}.
\end{corollary}
\begin{corollary}
    Let $\eta_{Cor:\ref{Cor:max_GCS}}$ denotes the number of GCSs that can be generated by Corollary \ref{Cor:max_GCS}. Then value of $\eta_{Cor:\ref{Cor:max_GCS}}$ is as follow:
    \begin{equation}
\eta_{Cor:\ref{Cor:max_GCS}}=\frac{p!}{2} \times (n!)^{2p-2} m^{n(p+1)}.
    \end{equation}
\end{corollary}
Due to the random choice of permutations ($\eta_i,\pi$) and arbitrary EBFs ($\phi_l,g$), the number of GCSs obtained by the proposed methods is more than the existing construction in the literature.
\begin{example}
    Let $n=m=2$, $p=2$, $k=3$, $\pi=(1,0)$, $\sigma_1=(2,0,1)$, $\sigma_2=(1,2,0)$, $\phi_0(x)=x$, $\phi_1(x)=x+1$, and $g(s_0,s_1,s_2)=s_0s_1+s_1s_2+s_0$ in Theorem \ref{Th:CCC}. Then we have
    \begin{equation}\label{Eq:frs_ex}
f^s_r=x_1x_0+r_0s_2+r_1s_0+r_2x_1+s_1x_0+x_0+x_1+1+s_0s_1+s_1s_2+s_0.
    \end{equation}
Corresponding to $f^s_r$, we have a binary $(8,4)$-CCC, which is given in Table \ref{tab:ex_CCC}.

\begin{table}[!htbp]
    \centering
    \small
    \begin{tabular}{|c|c|c|c|}
    \hline
       $C_0$  & $C_1$ & $C_2$ & $C_3$ \\
       \hline
       $\left[\begin{array}{rrrr}
-1 & 1 & 1 & 1 \\[-.5ex]
1 & -1 & -1 & -1 \\[-.5ex]
-1 & -1 & 1 & -1 \\[-.5ex]
-1 & -1 & 1 & -1 \\[-.5ex]
-1 & 1 & 1 & 1 \\[-.5ex]
1 & -1 & -1 & -1 \\[-.5ex]
1 & 1 & -1 & 1 \\[-.5ex]
1 & 1 & -1 & 1
\end{array}\right]$  & 
$\left[\begin{array}{rrrr}
-1 & 1 & 1 & 1 \\[-.5ex]
1 & -1 & -1 & -1 \\[-.5ex]
-1 & -1 & 1 & -1 \\[-.5ex]
-1 & -1 & 1 & -1 \\[-.5ex]
1 & -1 & -1 & -1 \\[-.5ex]
-1 & 1 & 1 & 1 \\[-.5ex]
-1 & -1 & 1 & -1 \\[-.5ex]
-1 & -1 & 1 & -1
\end{array}\right]$ &
$\left[\begin{array}{rrrr}
-1 & 1 & 1 & 1 \\[-.5ex]
-1 & 1 & 1 & 1 \\[-.5ex]
-1 & -1 & 1 & -1 \\[-.5ex]
1 & 1 & -1 & 1 \\[-.5ex]
-1 & 1 & 1 & 1 \\[-.5ex]
-1 & 1 & 1 & 1 \\[-.5ex]
1 & 1 & -1 & 1 \\[-.5ex]
-1 & -1 & 1 & -1
\end{array}\right]$ &
$\left[\begin{array}{rrrr}
-1 & 1 & 1 & 1 \\[-.5ex]
-1 & 1 & 1 & 1 \\[-.5ex]
-1 & -1 & 1 & -1 \\[-.5ex]
1 & 1 & -1 & 1 \\[-.5ex]
1 & -1 & -1 & -1 \\[-.5ex]
1 & -1 & -1 & -1 \\[-.5ex]
-1 & -1 & 1 & -1 \\[-.5ex]
1 & 1 & -1 & 1
\end{array}\right]$ \\
\hline
$C_4$ & $C_5$ & $C_6$ & $C_7$ \\
\hline 
$\left[\begin{array}{rrrr}
-1 & 1 & -1 & -1 \\[-.5ex]
1 & -1 & 1 & 1 \\[-.5ex]
-1 & -1 & -1 & 1 \\[-.5ex]
-1 & -1 & -1 & 1 \\[-.5ex]
-1 & 1 & -1 & -1 \\[-.5ex]
1 & -1 & 1 & 1 \\[-.5ex]
1 & 1 & 1 & -1 \\[-.5ex]
1 & 1 & 1 & -1
\end{array}\right]$ &
$\left[\begin{array}{rrrr}
-1 & 1 & -1 & -1 \\[-.5ex]
1 & -1 & 1 & 1 \\[-.5ex]
-1 & -1 & -1 & 1 \\[-.5ex]
-1 & -1 & -1 & 1 \\[-.5ex]
1 & -1 & 1 & 1 \\[-.5ex]
-1 & 1 & -1 & -1 \\[-.5ex]
-1 & -1 & -1 & 1 \\[-.5ex]
-1 & -1 & -1 & 1
\end{array}\right]$ &
$\left[\begin{array}{rrrr}
-1 & 1 & -1 & -1 \\[-.5ex]
-1 & 1 & -1 & -1 \\[-.5ex]
-1 & -1 & -1 & 1 \\[-.5ex]
1 & 1 & 1 & -1 \\[-.5ex]
-1 & 1 & -1 & -1 \\[-.5ex]
-1 & 1 & -1 & -1 \\[-.5ex]
1 & 1 & 1 & -1 \\[-.5ex]
-1 & -1 & -1 & 1
\end{array}\right]$ &
$\left[\begin{array}{rrrr}
-1 & 1 & -1 & -1 \\[-.5ex]
-1 & 1 & -1 & -1 \\[-.5ex]
-1 & -1 & -1 & 1 \\[-.5ex]
1 & 1 & 1 & -1 \\[-.5ex]
1 & -1 & 1 & 1 \\[-.5ex]
1 & -1 & 1 & 1 \\[-.5ex]
-1 & -1 & -1 & 1 \\[-.5ex]
1 & 1 & 1 & -1
\end{array}\right]$\\
\hline
    \end{tabular}
    \caption{$(8,4)$-CCC obtained from \eqref{Eq:frs_ex}}
    \label{tab:ex_CCC}
\end{table}
\end{example}

\begin{example}
    Let $n=m=6$, $\lambda=1$ $p=2$, $\pi=(0,1)$, $k=3$, $\sigma_1=(1,0,2)$, $\sigma_2=(0,1,2)$, $\phi_0(x)=2$ for $x\le 3$, $\phi_0(x)=x^2+3$ for $x\ge 4$, $\phi_1=x^3+4x+1$ and $g(s_0,s_1,s_2)=3s_0^2s_1+5s_1^2$ in Theorem \ref{Th:CCC}, we have
    \begin{align}
               f^s_r&=x_0x_1+r_0s_1+ r_1x_0+s_0x_1+r_2s_2+x_1^3+4x_1+1+3s_0^2s_1+5s_1^2+\left\{\begin{array}{ll}
		2,& \text{for } x_0\le 3, \\
		x_0^2+3, & \text{for } x_0\ge 4, 
	\end{array}\right.\nonumber \\
 &=\left\{\begin{array}{ll}
		x_0x_1+r_0s_1+ r_1x_0+s_0x_1+r_2s_2+x_1^3+4x_1+3s_0^2s_1+5s_1^2+3,&\text{for } x_0\le 3, \\
		x_0x_1+r_0s_1+ r_1x_0+s_0x_1+r_2s_2+x_1^3+4x_1+3s_0^2s_1+5s_1^2+x_0^2+4, &\text{for } x_0\ge 4.
	\end{array}\right. \nonumber
    \end{align}
    Using this function $ f_r^s$, we construct a $(216, 36)$-CCC, which are not available in literature.



\end{example}
\section{Comparison with existing literature of CCCs}
The proposed method yield a larger number of distinct CSs and CCCs with the same parameters, as the proposed construction employs a more general functional form than the constructions in \cite{2008_rathinakumar, sarkar2024investigation,Shen_CCC}. Additionally, the proposed construction supports CCCs with additional parameters not previously available. In our construction, the EBFs $\phi_l$'s are not restricted to polynomial, they may be any map from $\mathbb{Z}_m \to \mathbb{Z}_m$, including non-polynomial functions such as piecewise maps.

The proposed construction generates a CCC where the number of codes exceeds the code length, an aspect missing from the existing literature. Further, the PMEPR of each sequence in our construction is bounded by $n$, whereas in existing literature the PMEPR is bounded by $n^k$ (the set size). A detailed comparison with existing methods is provided in Table~\ref{tab:comparison}.
\begin{table}[htbp] \small
\centering
     {
\begin{tabular}{|c|c|c|c|c|c|c|}
\hline
\textbf{Ref.} & $(M,N)$\textbf{-CCC}  & \textbf{$M>N$} & \textbf{Constraints} &  \textbf{column} & \textbf{row} \\ 
 &   &  &  &  \textbf{PAPR}&  \textbf{PAPR} \\
&   & &  &  \textbf{upper}&  \textbf{upper}  \\
 &   & &  &  \textbf{bound} &  \textbf{bound} \\
\hline
\cite{2008_rathinakumar} & $\left(2^k, 2^m\right)$  & No & $m, k \in \mathbb{Z}^{+}, m\geq k$ & $2^{k}$ & - \\ 
\hline
{\cite{2023_kumar}} & $\left(2^{k},  10\cdot2^m\right),$  & No & $m, k \in \mathbb{Z}^{+}, m \geq k$ &$2$ & - \\ 
& $\left(2^{k},  26\cdot2^m\right)$ &  &  &  & \\ 
\hline
\cite{Shen_CCC} & $(n^k,  n^p)$  & No & $n \geq 2,  k \geq p \geq 1$ & $n^k$ & - \\
\hline
\cite{sarkar2024investigation} & 
$\left(\prod_{i=1}^{k} p_i^{n_i}, \prod_{i=1}^k p_i^{m_i}\right)$
  & No & $ m_i, n_i \in \mathbb{Z}^{+}, n_i \leq m_i$, &  $\prod_{i=1}^{k} p_i^{n_i}$&  - \\ 
  &   & & $p_i$ is prime, $1 \leq i \leq k$  & &  \\
  \hline 
\cite{204_liu_8} & $\left(2^k, 2^m\right)$  & No & $m, k \in \mathbb{Z}^{+},  m\geq k$ & $2$ & - \\ 
\hline

\cite{2008_chen_ccc} & $\left(2^k,  2^m\right)$   & No & $k, m \in \mathbb{Z}^{+}, k \leq m$ & $2^k$ & - \\ 
\hline
\cite{2020_tian_31} & $\left(2^{k}, 2^m\right)$  & No & $m, k \in \mathbb{Z}^{+}, m\geq k$ & $2^{k}$ & - \\ 
\hline
\cite{2008_wu_32} & $\left(2^k, 2^m\right)$   & No & $m, k \in \mathbb{Z}^{+}, k \leq m$  & $2^k$ & - \\ 
\hline
\cite{2018_das_41} & $(n,n^n)$  & No & $1\ge n\in \mathbb{Z}^+$  & $n$ & - \\ 
\hline
\cite{2018_das_42} & $(M,  P^m)$   & Yes & $m \geq 1, P \mid M$  & $M$ & - \\ 

\hline
\cite{2019_ma_45} & $(2^k,  2^{km})$   & No & $ m, k \in \mathbb{Z}^{+}$ & - & - \\ 
\hline

 Th. 1 & $(n^k,  n^p)$  & Yes & $n \geq 2,  p, k \in \mathbb{Z}^{+}$ & $n^k$ & $n$ \\
 Cor. \ref{Cor:Col_PMEPR_red} & $(n^k,  n^p)$  & Yes &  & $n$ & $n$ \\
\hline
\end{tabular}}
\caption{Comparative Analysis of the Proposed  Construction with  existing CCCs.}
\label{tab:comparison}
\end{table}

\section{Conclusion}

This paper presented a novel construction of CCC based on  EBFs and permutation.  The key advantage of the proposed approach is its ability to generate a CCC with a number of codes that significantly exceeds the code length, an aspect not previously explored in the literature. This property enables support for a larger number of users in communication systems. Further, the PMEPR of each sequence in the proposed construction is bounded by  $n$, which is substantially lower than the PMEPR constraint of $n^k$ in existing methods. 
 \bibliographystyle{IEEEtran}
 \bibliography{Rajen}
\end{document}